\documentclass[preprint,12pt]{elsarticle}
\pdfoutput=1

\usepackage{epsfig}
\usepackage{float}
\usepackage{graphicx}
\usepackage{times}
\usepackage{amsmath,amsthm}

\biboptions{numbers,sort&compress}

\usepackage{amssymb}

\journal{Proceedings of the Royal Society A-Mathematical Physical and Engineering Sciences}

\begin{document}

\begin{frontmatter}

\title{Indirect exclusion can promote cooperation in repeated group interactions}

\author[label1,label2]{Linjie Liu}

\author[label2]{Xiaojie Chen\corref{cor}}
\cortext[cor]{Corresponding author} \ead{xiaojiechen@uestc.edu.cn}

\address[label1]{College of Science, Northwest A \& F University, Yangling 712100, China}
\address[label2]{ School of Mathematical Sciences, University of Electronic Science and Technology of China, Chengdu 611731, China}


\begin{abstract}
Social exclusion has been regarded as one of the most effective measures to promote the evolution
of cooperation. In real society, the way in which social exclusion works can be direct or indirect.
However, thus far there is no related work to explore how indirect exclusion influences the evolution of
cooperation from a theoretical perspective. Here, we introduce indirect exclusion into the repeated
public goods game where the game organizer probabilistically selects cooperators after the
first game round to participate in the following possible game interactions. We then investigate
the evolutionary dynamics of cooperation both in infinite and finite well-mixed populations. Through
theoretical analysis and numerical calculations, we find that the introduction of indirect exclusion
can induce the stable coexistence of cooperators and defectors or the dominance of cooperators,
which thus effectively promotes the evolution of cooperation. Besides, we show that the identifying
probability of the organizer has a nonlinear effect on public cooperation when its value is lower than
an intermediate value, while the higher identifying probability can maintain a high level of cooperation.
Furthermore, our results show that increasing the average rounds of game interactions can effectively promote the evolution of cooperation.
\end{abstract}


\end{frontmatter}

\newtheorem{thm}{\bf Theorem}[section]
\newtheorem{remark}{Remark}[section] 

\section{Introduction}
Cooperation is crucial to the maintenance and stability of life systems at all levels, from microscopic cells to extremely complex biosphere \cite{Perc_PR_17,Perc_bio_10,Perc_jrsi_13,Perc_NJP_11,Szolnoki2014interface,Ito18,Tanimoto18}. For example, bacterial strains that provide each other with missing essential amino acids (cooperative cross-feeding interactions) have higher fitness compared with metabolic autonomy \cite{Pande_isme_2014}. Bats usually cooperate in roosting, foraging, feeding  or caring for their offspring \cite{Wilkinson_prsb_2016}. In human society, climate change \cite{Vasconcelos_ncc_2013,Vasconcelos_pnas_2014}, corruption \cite{Lee_pnas_2019,liu_mmmas_2019,liu_amc_2022}, and the spread of diseases \cite{Karlsson_sr_2020} are huge challenges that remind us that global cooperation is necessary. Although cooperation is socially desirable and widespread in nature and human society, it often incurs individual own costs to bring benefits to other individuals, resulting in that cooperation is not favored by natural selection \cite{Axelrod1984,Szolnoki2012pre,Szolnoki2022csf,han18sr,Duong2021PRSA,li21csf,zhu22ITNE,zhu20ITCS,xia2016ISJ}. Therefore, it has always been a great challenge to explain how cooperative behaviour evolves \cite{Kennedy_science_2005}.

Several mechanisms have been proposed to explain the evolution of cooperation, among which
the role of incentives in promoting cooperation has been widely studied by researchers in a wide
range of fields  \cite{perc2012njp,perc2012sr,Szolnoki2015prsb,chen2015pre,Chen_15sr,Szolnoki2017,wang_cnsns_2019,sun_is_2021,dong_prsb_2019,hu_csf_2020,Szolnoki_epl_20100,Han2021P,han2022risf,fang2019prsa,wang2021prsa}. Previous experiments and theoretical studies have revealed that highlevel cooperation can be effectively maintained by excluding free-riders from the beneficiaries \cite{sasaki_prsb_2013,li_pre_2015,li_epl_2016,szolnoki_pre_2017,liu_ND_2019,quan_physica_2019,quan_chaos_2019,zhao_BMC_2019,quan_amc_2020,quan_csf_2021,liu_kbs_2020,liu_jrsf_2022}. Concretely, Sasaki and Uchida introduced social exclusion strategy into the public goods
game (PGG) and studied its evolutionary dynamics in infinite populations \cite{sasaki_prsb_2013}. Their theoretical results show that social exclusion can better promote the evolution of cooperation than costly punishment. Li \emph{et al.} compared the effects of peer exclusion and pool exclusion on resisting defectors in the presence of behavioural mutation and decision-making error \cite{li_pre_2015}. They revealed that peer exclusion can do better than pool exclusion when exclusion cost is small. Recently, the evolutionary dynamics of social exclusion strategy have also been explored in the framework of structural populations \cite{szolnoki_pre_2017,quan_chaos_2019,quan_csf_2021} and repeated group interactions \cite{liu_jrsf_2022}.

Nevertheless, in most of the above theoretical game models, it is often assumed that social
exclusion can be used to solve the free-riding problem in a direct way. In other words, excluders
pay the cost to expel free-riders in game interactions when free-riders are identified. In the
process of daily interactions, exclusion may also appear in an indirect way without cost, that
is, purposefully avoiding free-riders from participating in game interactions \cite{Lindst_NHB_2018,liu_epl_2013}. For example,
an experimental study has revealed that chimpanzees can remember the cooperative performance
of their partners in previous interactions and then tend to choose more cooperative partners in
subsequent interactions, resulting in fewer opportunities for the less effective partner to cooperate \cite{Melis_science_2006}. Such recruitment activities occur more frequently when they need assistance to obtain
the food. In addition, individuals can avoid interactions with cheaters by partner switching
mechanism, which has been proved to stabilize cooperation in cleaning fish \cite{Bshary_BL_2005} and human society \cite{fu_pre_08,fu_pre_09,Du_DGA_11}. However, these existing studies characterizing the indirect form of exclusion are
based on computer simulations and behavioural experiments \cite{Melis_science_2006,fu_pre_09,liu_epl_2013}.On the other hand, in the
process of repeated interactions, there may be errors in memorizing or recognizing the strategies
of other individuals \cite{Reiter_18,Hilbe_18}, especially in searching for cooperative partners, which may lead the
decision-maker to choose the wrong partners. To our knowledge, thus far few theoretical works
have studied how such an indirect exclusion approach influences the evolution of cooperation
when decision error exists. Particularly, it is still unclear how the introduction of this approach
will affect the evolution of cooperation from a theoretical perspective.

To fill this gap, we introduce indirect exclusion into the repeated PGG and explore its
evolutionary dynamics both in infinite and finite populations. We consider the social selection
process, in which the game organizer probabilistically selects cooperators after the first game
round and then organizes them to participate in the following possible game interactions, and
thus those who are not selected will lose the opportunity to participate in game interactions. Our
theoretical and numerical results show that the introduction of indirect exclusion can effectively
promote the evolution of cooperation. Particularly, we find that longer game rounds and larger
identifying probability can better maintain a high level of cooperation.

\section{Model and Methods}

We consider the PGG played in a well-mixed population where $N$ individuals are randomly selected to form a game group. At each game round, cooperators ($C$) contribute $c$ to the common pool, while defectors ($D$) contribute nothing. The sum of contributions is
multiplied by a synergy factor $F$ where $1<F<N$ and then equally allocated among all group members. The above process repeats itself with the probability $w$ (also called discount factor, describing the probability that future payoffs may be discounted \cite{Hilbe_18nature}), and hence the expected number of game rounds is $T=\frac{1}{1-w}$. If no additional mechanism works, the payoff of cooperators obtained from the $T$ rounds game interactions is always lower than the payoff of defectors in a given group, resulting in that each individual prefers to defect in the population. 

In order to focus on the effect of indirect exclusion on the evolution of cooperation,  in repeated interactions we do not consider the strategy with memory or conditional strategy, which has been proved to influence the evolutionary dynamics of cooperation via direct reciprocity \cite{Hilbe_18}. Concretely, before the game interaction, one individual is randomly selected as the game organizer. After the first round, the organizer knowing the numbers of cooperators and defectors among the $N-1$ co-players, is responsible for selecting all cooperators from the first game round and organize the subsequent possible game interactions. We assume that the probability of the organizer successfully selecting one cooperator is $p$. In this way, those individuals who are not selected will be excluded from participating in the subsequent game interactions and can only get the benefits from the first game round, and the accumulated contributions of the subsequent game interactions will be allocated equally to the selected individuals. 
Accordingly, the payoffs of cooperators and defectors from $T$ rounds consist of the following three parts:\\

(1) The payoffs of cooperators and defectors from the first game round can be given by
\begin{eqnarray}
   \pi_{C}^{(0)}&=&\frac{Fc(N_{C}+1)}{N}-c,\\
   \pi_{D}^{(0)}&=&\frac{FcN_{C}}{N},
\end{eqnarray}
where $N_{C}$ denotes the number of cooperators in the group.

(2) When the focal individual is the organizer who successfully selects $N_{OC}$ cooperators from the remaining $N-1$ individuals, then the payoffs of cooperators and defectors can be respectively written as
\begin{eqnarray}
   \pi_{C}^{(1)}&=&\frac{1}{N}[\frac{Fc(N_{OC}+1)}{N_{C}+1}-c](T-1),\\
   \pi_{D}^{(1)}&=&\frac{1}{N}\frac{FcN_{OC}}{N_{C}+1}(T-1),
\end{eqnarray}
where $\frac{1}{N}$ denotes the probability that the focal individual is the organizer.

(3) When the focal individual is not the organizer, then the payoffs of cooperators and defectors can be respectively given by
\begin{eqnarray}
   \pi_{C}^{(2)}&=&\frac{N-1}{N}[\frac{Fc(N_{OC}+1)}{N_{C}+1}-c]p(T-1),\\
   \pi_{D}^{(2)}&=&\frac{N-1}{N}\frac{FcN_{OC}}{N_{C}+1}(1-p)(T-1),
\end{eqnarray}
where $\frac{N-1}{N}$ denotes the probability that the focal individual is not the organizer.

Combining the aforementioned equations, we can respectively write the payoffs of cooperators and defectors in repeated group interactions as
\begin{eqnarray}\label{eq7}
   \pi_{C}&=&\pi_{C}^{(0)}+\pi_{C}^{(1)}+\pi_{C}^{(2)}\nonumber\\
&=&\frac{Fc(N_{C}+1)}{N}-c+[\frac{Fc(N_{OC}+1)}{N_{C}+1}-c]\frac{(T-1)[1+p(N-1)]}{N}
\end{eqnarray}
and
\begin{eqnarray}\label{eq8}
   \pi_{D}&=&\pi_{D}^{(0)}+\pi_{D}^{(1)}+\pi_{D}^{(2)}\nonumber\\
&=&\frac{FcN_{C}}{N}+\frac{(T-1)[1+(1-p)(N-1)]}{N}\frac{FcN_{OC}}{N_{C}+1}.
\end{eqnarray}

In the framework of evolutionary game theory, individuals tend to imitate strategies of others when these can produce higher payoffs. In the following, we shall give our methods to investigate the evolutionary dynamics of cooperation and defection in infinite and finite well-mixed populations.

\subsection{Replicator equation}

In an infinite well-mixed population, we investigate the evolutionary dynamics of cooperation and defection strategies by analysing the replicator equation \cite{Schuster_83,Hofbauer_98}, which describes the change in the cooperation level in the population. Concretely, the replicator equation is given by
\begin{eqnarray}\label{re}
\dot{x}=x(1-x)(P_{C}-P_{D}),
\end{eqnarray}
where $P_{C}$ and $P_{D}$ are the average payoffs of $C$ and $D$ individuals. According to Eq. (\ref{re}), we know that when $P_{C}>P_{D}$, then $\dot{x}>0$, which means that the frequency of cooperators in the population will increase. The average payoffs of cooperators and defectors can be written as

\begin{eqnarray*}\label{equ8}
   P_{C}&=&\sum_{N_{C}=0}^{N-1}\binom{N-1}{N_{C}}\sum_{N_{OC}=0}^{N_{C}}\binom{N_{C}}{N_{OC}}p^{N_{OC}}(1-p)^{N_{C}-N_{OC}}x^{N_{C}}(1-x)^{N-N_{C}-1}\pi_{C},\nonumber\\
   P_{D}&=&\sum_{N_{C}=0}^{N-1}\binom{N-1}{N_{C}}\sum_{N_{OC}=0}^{N_{C}}\binom{N_{C}}{N_{OC}}p^{N_{OC}}(1-p)^{N_{C}-N_{OC}}x^{N_{C}}(1-x)^{N-N_{C}-1}\pi_{D},
\end{eqnarray*}

where $\pi_{C}$ and $\pi_{D}$ are shown in equations (\ref{eq7}) and (\ref{eq8}), respectively.

\subsection{Stochastic dynamics}

However, when the population size is finite, the evolutionary dynamics will be affected by errors
of imitation and behavioural mutations, and thus replicator equations of the previous section
cannot be used to describe the stochastic dynamics in finite populations. For finite well-mixed
populations of size $Z$, with $i_{C}$ cooperators and $Z-i_{C}$ defectors, the average payoffs of cooperators and defectors are now respectively given by
\setlength{\arraycolsep}{0.0em}
\begin{eqnarray}
f_{C}&=&\sum_{N_{C}=0}^{N-1}\frac{\binom{i_{C}-1}{N_{C}}\binom{Z-i_{C}}{N-N_{C}-1}}{\binom{Z-1}{N-1}}\sum_{N_{OC}=0}^{N_{C}}\binom{N_{C}}{N_{OC}}p^{N_{OC}}(1-p)^{N_{C}-N_{OC}}\pi_{C}
\end{eqnarray}
\setlength{\arraycolsep}{5pt}
and
\setlength{\arraycolsep}{0.0em}
\begin{eqnarray}
f_{D}&=&\sum_{N_{C}=0}^{N-1}\frac{\binom{i_{C}}{N_{C}}\binom{Z-i_{C}-1}{N-N_{C}-1}}{\binom{Z-1}{N-1}}\sum_{N_{OC}=0}^{N_{C}}\binom{N_{C}}{N_{OC}}p^{N_{OC}}(1-p)^{N_{C}-N_{OC}}\pi_{D},
\end{eqnarray}
\setlength{\arraycolsep}{5pt}
where $\pi_{C}$ and $\pi_{D}$ are shown in equations (\ref{eq7}) and (\ref{eq8}), respectively.

Now, we adopt the pairwise comparison rule to study the evolutionary dynamics of
cooperation and defection in the finite population. Concretely, the probability that individual $A$
adopts the strategy of a randomly selected individual $B$ is given by the Fermi function \cite{szabo1998evolutionary}
\begin{eqnarray}
p_{AB}=\frac{1}{1+e^{-s(f_{B}-f_{A})}},
\end{eqnarray}
where $s$ is the intensity of selection that determines the level of uncertainty in the strategy
adoption process.
For $s\rightarrow 0$, the selection is weak and an individual imitates the strategy of others
randomly. Whereas for $s\rightarrow \infty$, a more successful player is always imitated, which is regarded as
a strong imitation.

With these descriptions, the probability that the number of $C$ individuals in the population increases or decreases by one is given by
\begin{eqnarray}\label{eq15}
T^{\pm}(i_{C})=\frac{i_{C}}{Z}\frac{Z-i_{C}}{Z}\frac{1}{1+e^{\mp s(f_{C}-f_{D})}}.
\end{eqnarray}

The gradient of selection, which is described by equation (\ref{re}) for infinite populations, is replaced by the following equation for finite populations
\begin{eqnarray}
G(i_{C})\equiv T^{+}(i_{C})-T^{-}(i_{C})=\frac{i_{C}}{Z}\frac{Z-i_{C}}{Z}\tanh[\frac{s}{2}(f_{C}-f_{D})].
\end{eqnarray}

We investigate the evolutionary dynamics of cooperation in finite populations by analysing
the gradient of selection in the absence of mutations. When considering behavioural mutations,
one alternative and appropriate approach to study the evolutionary dynamics in a finiteIt is worth noting
that the existence of behavioural mutation induces that the population will never fixate in
monomorphic states, and the stationary distribution provides information about the time spent
by the population on each configuration. Here, we use $T_{\textbf{i}(i_{CI})\rightarrow \textbf{i}^{'}(i_{CI}^{'})}$ to describe the probabilities of system transition from state $\textbf{i}$ to state $\textbf{i}^{'}$, which are defined as
\begin{equation}
T_{\textbf{i}(i_{CI})\rightarrow \textbf{i}^{'}(i_{CI}^{'})}=\left\{
\begin{array}{rcl}
0 & & {\text{if} \quad |i_{CI}^{'} - i_{CI}|>1}\\
T_{\mu}^{+}(i_{C}) & & {\text{if} \quad i_{CI}^{'} - i_{CI}=1}\\
T_{\mu}^{-}(i_{C}) & & {\text{if} \quad i_{CI} - i_{CI}^{'}=1}\\
1-T_{\mu}^{+}(i_{C})-T_{\mu}^{-}(i_{C}) & & \text{otherwise},
\end{array} \right.
\end{equation}
where $T_{\mu}^{+}(i_{C}) =(1-\mu)T^{+}(i_{C})+\mu(Z-i_{C})/Z$ and $T_{\mu}^{-}(i_{C}) =(1-\mu)T^{-}(i_{C})+\mu i_{C}/Z$, respectively, denote the probabilities that the number of cooperators in the population increases and decreases one. $T^{\pm}(i_{C})$ is defined in equation (\ref{eq15}) and $\mu$ is the mutation rate.
By characterizing the transition probabilities between any two states, we can obtain the tridiagonal transition matrix $W=[T_{\textbf{i}\textbf{i}^{'}}]_{(Z+1)\times (Z+1)}^{T}$. Then the stationary distribution $\textbf{P}$ of the Markov process can be obtained from the eigenvector corresponding to the eigenvalue 1 of $W$ \cite{VanKampen1992}.

Furthermore, we provide an indicator $l_{C}$ to measure the average level of cooperation, which is given as follows,
\begin{eqnarray}
l_{C}=\frac{\textbf{S}\textbf{P}}{Z},
\end{eqnarray}
where the vector $\textbf{S}=[0,\cdots,Z]$ denotes the population states.

\section{Results}

\subsection{Evolutionary dynamics in infinite well-mixed populations}

We first study the evolutionary dynamics of cooperation induced by indirect exclusion in an infinite well-mixed population by analysing the replicator equation (\ref{re}). By calculation, we can respectively write the average payoffs of these two strategies as
\begin{eqnarray*}
   P_{C}&=&\frac{Fc(N-1)x}{N}+\frac{Fc}{N}-c+\frac{c[1+(N-1)p](T-1)}{N^{2}x}\big\{F-F(1-x)^{N}\\
&-&Nx+Fp[(1-x)^{N}-1+Nx]\big\},\\
   P_{D}&=&\frac{Fc(N-1)x}{N}+\frac{Fc[p+N(1-p)]p(T-1)[Nx-1+(1-x)^{N}]}{N^{2}x}.
\end{eqnarray*}

The distribution and stability of the equilibrium points of the replicator equation (\ref{re}) can be given by the following theorem. For analytical convenience, we denote that $G(x)=x(1-x)Q(x)$, where $Q(x)=P_{C}-P_{D}$.

\begin{thm}\label{T0.1}
Let $\theta_{1}=[2(N-1)Fp(1-p)+F]+(1+Np-p)(Fp-1)-F(p+N-Np)p$ and $\theta_{2}=[2(N-1)Fp(1-p)+F]+(1+Np-p)N(Fp-1)-F(p+N-Np)Np,$ then we have the following conclusions:\\
(1) When $Q(0)<0$, that is, $(T-1)\theta_{1}<N-F$, the replicator equation (\ref{re}) has no interior equilibrium point. The boundary equilibrium point $x=0$ is stable, while $x=1$ is unstable.\\
(2) When $Q(1)<0<Q(0)$, that is, $\frac{(T-1)\theta_{2}}{N}<N-F<(T-1)\theta_{1}$, the replicator equation (\ref{re}) has an interior equilibrium point, which is stable. The two boundary equilibrium point $x=0$ and $x=1$ are both unstable.\\
(3) When $Q(1)>0$, that is, $\frac{(T-1)\theta_{2}}{N}>N-F$, the replicator equation (\ref{re}) has no interior equilibrium point. The boundary equilibrium  point $x=0$ is unstable, while $x=1$ is stable.
\end{thm}

\begin{figure}[!t]
\centering\includegraphics[width=5in]{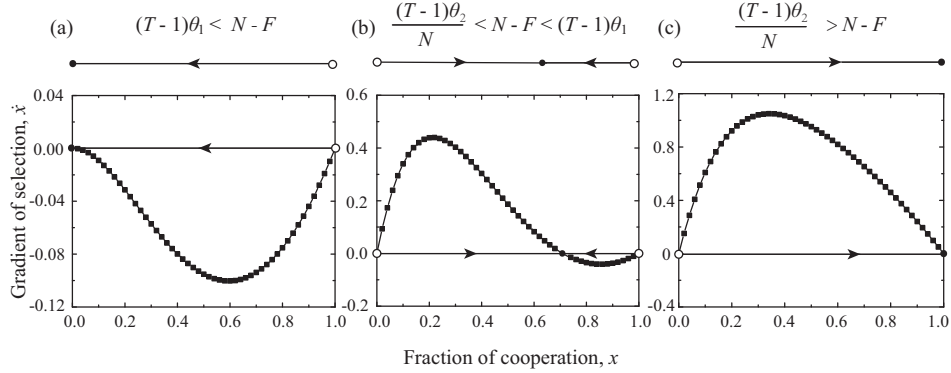}
\caption{The gradient of selection $\dot{x}$ changes with the fraction of cooperators for different model parameters. Open circles denote unstable equilibria, and solid points represent stable equilibria. The arrow pointing to the right indicates that cooperators are more favored than defectors. Parameter values: $N=5, c=1, F=3, T=2,$ and $p=0.01$ in panel (a); $N=5, c=1, F=3, T=5,$ and $p=0.6$ in panel (b); $N=5, c=1, F=3, T=6,$ and $p=0.8$ in panel (c).}
\label{fig1}
\end{figure}

\begin{proof}
According to equation (\ref{re}), we know that $x=0$ and $x=1$ are two boundary equilibrium points of the system. The existence of interior equilibrium point is determined by $Q(x)$. Considering that $1-(1-x)^{N}=x\sum_{k=0}^{N-1}(1-x)^{k}$, we can calculate the difference between $P_{C}$ and $P_{D}$ as
\begin{eqnarray}\label{equ12}
Q(x)&=&\frac{Fc}{N}-c+\frac{c(T-1)}{N^{2}}\big\{[2(N-1)Fp(1-p)+F]\sum_{k=0}^{N-1}(1-x)^{k}\nonumber\\
&+&(1+Np-p)N(Fp-1)-F(p+N-Np)Np\big\}.
\end{eqnarray}
We can get 
\begin{eqnarray}\label{equ13}
Q'(x)&=&-\frac{Fc(T-1)}{N^{2}}[2(N-1)p(1-p)+1]\sum_{k=1}^{N-1}k(1-x)^{k-1}.
\end{eqnarray}
Obviously, $Q'(x)<0$ for $x\in(0,1)$. Therefore, the value of $Q(x)$ decreases monotonically with the increase of $x$.
Besides, we have 
\begin{eqnarray*}
Q(0)&=&\frac{Fc}{N}-c+\frac{c(T-1)}{N}\big\{[2(N-1)Fp(1-p)+F]\nonumber\\
&+&(1+Np-p)(Fp-1)-F(p+N-Np)p\big\},
\end{eqnarray*}
\begin{eqnarray*}
Q(1)&=&\frac{Fc}{N}-c+\frac{c(T-1)}{N^{2}}\big\{[2(N-1)Fp(1-p)+F]\nonumber\\
&+&(1+Np-p)N(Fp-1)-F(p+N-Np)Np\big\}.
\end{eqnarray*}

Thus we can get the following conclusions:\\
(1) When $Q(0)<0$, which means $(T-1)\theta_{1}<N-F$, then the gradient of selection $\dot{x}$ is always negative for all $x \in (0,1)$. Thus Eq. (\ref{re}) has no interior equilibrium point. In this case, we can judge $G'(0)=Q(0)<0$ and $G'(1)=-Q(1)>0$. Therefore, $x=0$ is a stable equilibrium point, while $x=1$ is an unstable equilibrium point.\\
(2) When $Q(1)<0<Q(0)$, that is, $\frac{(T-1)\theta_{2}}{N}<N-F<(T-1)\theta_{1}$, equation (\ref{re}) has an interior equilibrium point $x=x^{*}$. Because $G'(x^{*})=x^{*}(1-x^{*})Q'(x^{*})<0$, we can judge that the interior equilibrium point is stable. Besides, since $G'(0)=Q(0)>0$ and $G'(1)=-Q(1)>0$, we know that $x=0$ and $x=1$ are both unstable.\\
(3) When $Q(1)>0$, that is, $\frac{(T-1)\theta_{2}}{N}>N-F$, then the gradient of selection $\dot{x}$ is always positive for all $x \in (0,1)$. Thus equation (\ref{re}) has no interior equilibrium point. Since $G'(0)=Q(0)>0$ and $G'(1)=-Q(1)<0$, we know that $x=0$ is an unstable equilibrium point and $x=1$ is stable.
\end{proof}

We provide some numerical calculations to verify the above theoretical analysis, as shown in figure~\ref{fig1}. When $(T-1)\theta_{1}<N-F$, we find the values of gradient of selection $\dot{x}$ are negative for all $x\in(0,1)$. The arrow points to the left, which means that defection is the strategy which has evolutionary advantage (see figure~\ref{fig1}(a)). When $\frac{(T-1)\theta_{2}}{N}<N-F<(T-1)\theta_{1}$, a stable interior equilibrium point can appear, and thus cooperators and defectors can coexist steady in the population (see figure~\ref{fig1}(b)). Particularly, when $\frac{(T-1)\theta_{2}}{N}>N-F$, the values of gradient of selection $\dot{x}$ are always positive for all $x\in(0,1)$ and the arrow points to the right, which means that cooperation is favored over defection (see figure~\ref{fig1}(c)).

\begin{figure}[t]
\centering
\centering\includegraphics[width=5in]{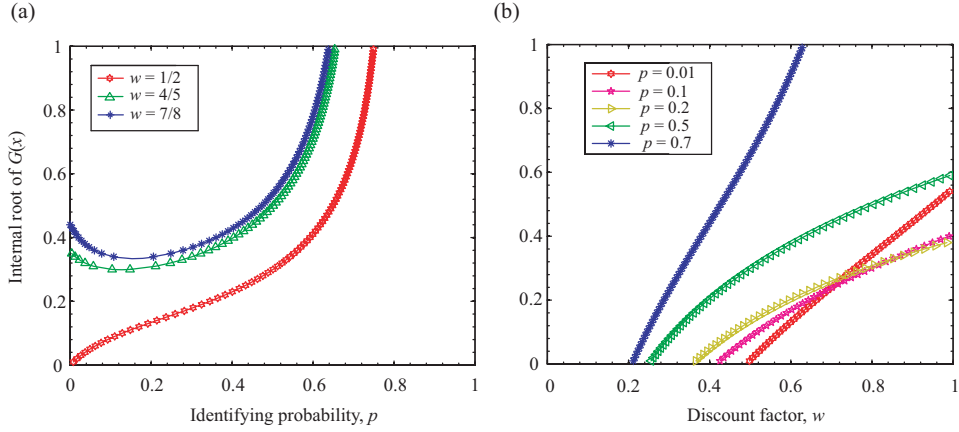}
\caption{The internal root of $G(x)$ changes with different parameters in infinite populations. Panel (a) shows the internal roots of $G(x)$ as a function of $p$ for different values of $w$. Panel (b) shows the internal roots of $G(x)$ as a function of $w$ for different values of $p$. Parameter values: $N=5, c=1$, and $F=3$ in panels (a) and (b).}
\label{fig2}
\end{figure}

Note that the internal root of $G(x)$ determines the cooperation level at equilibrium when cooperators can coexist with defectors in the population. In figure~\ref{fig2}, we show how the internal root of $G(x)$ changes with model parameters $p$ and $w$, which are two important parameters affecting the stable cooperation level. We find that when the identifying probability of the organiser $p$ is higher, there is no interior equilibrium point. When the model parameters satisfy $\frac{(T-1)\theta_{2}}{N}<N-F<(T-1)\theta_{1}$, a stable interior equilibrium point can appear. Particularly, when the discount factor is small ($w=1/2$), the value of the internal root increases with the increase of $p$ (see figure~\ref{fig2}(a)). When the value of $w$ is large, we find that the existing interior equilibrium point first decreases, reaches a minimum value, but then increases with the increase of $p$. Besides, a higher $p$ value can maintain a higher level of cooperation. In order to show the role of $w$ in evolutionary outcomes more clearly, we further investigate the effect of $w$ on the value of the existing interior equilibrium point. As shown in figure~\ref{fig2}(b), we find that the value of the existing interior equilibrium point increases with increasing $w$. Interestingly, when the value of $p$ is slightly small (for example, $p\leq 0.2$), with the increase of $w$, the larger the $p$ value, the slower the growth of the internal root value.

\subsection{Evolutionary dynamics in finite well-mixed populations}

\begin{figure}[t]
\centering
\centering\includegraphics[width=5in]{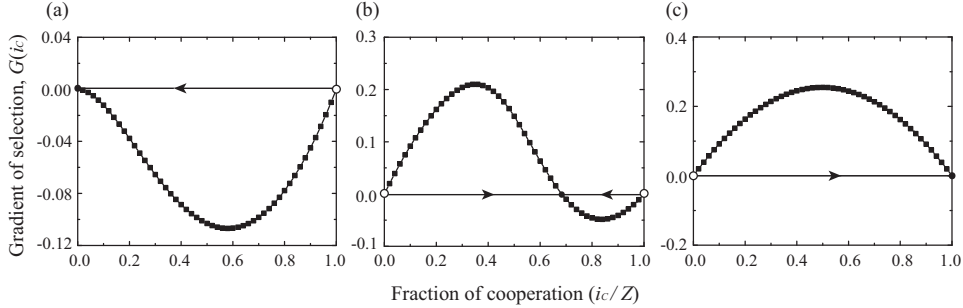}
\caption{The gradient of selection changes with the fraction of cooperators for different parameters in finite populations. Parameter values: $Z=50, N=5, c=1, F=3, \mu=0.01, s=2, T=2,$ and $p=0.01$ in panel (a); $Z=50, N=5, c=1, F=3, \mu=0.01, s=2, T=5,$ and $p=0.6$ in panel (b); $Z=50, N=5, c=1, F=3, \mu=0.01, s=2, T=6,$ and $p=0.8$ in panel (c).}
\label{fig3}
\end{figure}

In what follows, we study the evolutionary dynamics of indirect exclusion in a finite population, in which behaviour mutation and decision error are involved. In figure~\ref{fig3}, we investigate how the gradient of selection $G(i_{C})$ changes with the initial fraction of cooperators in different parameter regions in a finite population. We find that the three evolutionary outcomes showed in infinite populations
can still appear in a finite population. Figure~\ref{fig3}(a) shows that the gradient of selection is always negative and cooperation cannot be maintained in the population. Figure~\ref{fig3}(b) shows that there exists a stable interior equilibrium point, which means that cooperators can coexist with defectors in the population. The third typical outcome shown in figure~\ref{fig3}(c) reveals that the gradient of selection is always positive and full cooperation is the steady state of the system.

\begin{figure}[t]
\centering
\centering\includegraphics[width=5in]{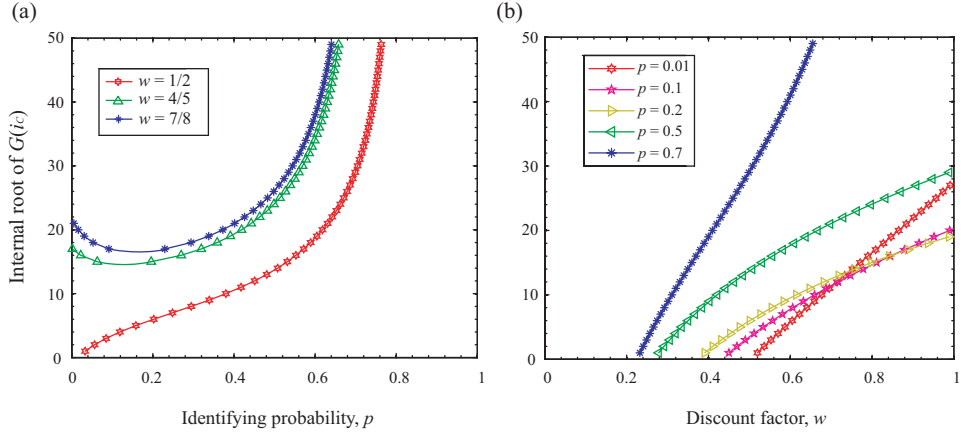}
\caption{Internal root of $G(i_{C})$ changes with model parameters in finite populations. Parameter values: $Z=50, N=5, c=1, \mu=0.01, s=2$, and $F=3$ in panels (a) and (b).}
\label{fig4}
\end{figure}

Similar to figure~\ref{fig2} for infinite populations, we show that the internal root of $G(i_{C})$ varies with
the identifying probability $p$ and the discount factor $w$ in the finite population. We find that when $w$ is small, the internal root of $G(i_{C})$ increases with the increase of $p$. When the value of $w$ is slightly high, the value of the internal root first decreases, reaches the minimum value, and then increases with the increase of $p$ (see Fig.~\ref{fig4}(a)). Furthermore, the value of the internal root monotonically increases
with increasing $w$ (see figure~\ref{fig4}(b)). The above outcomes indicate that a larger identifying probability or larger discount factor can better promote the emergence of cooperation, which are also found in the infinite population (see figure~\ref{fig2}).

\begin{figure}[t]
\centering
\includegraphics[width=5in]{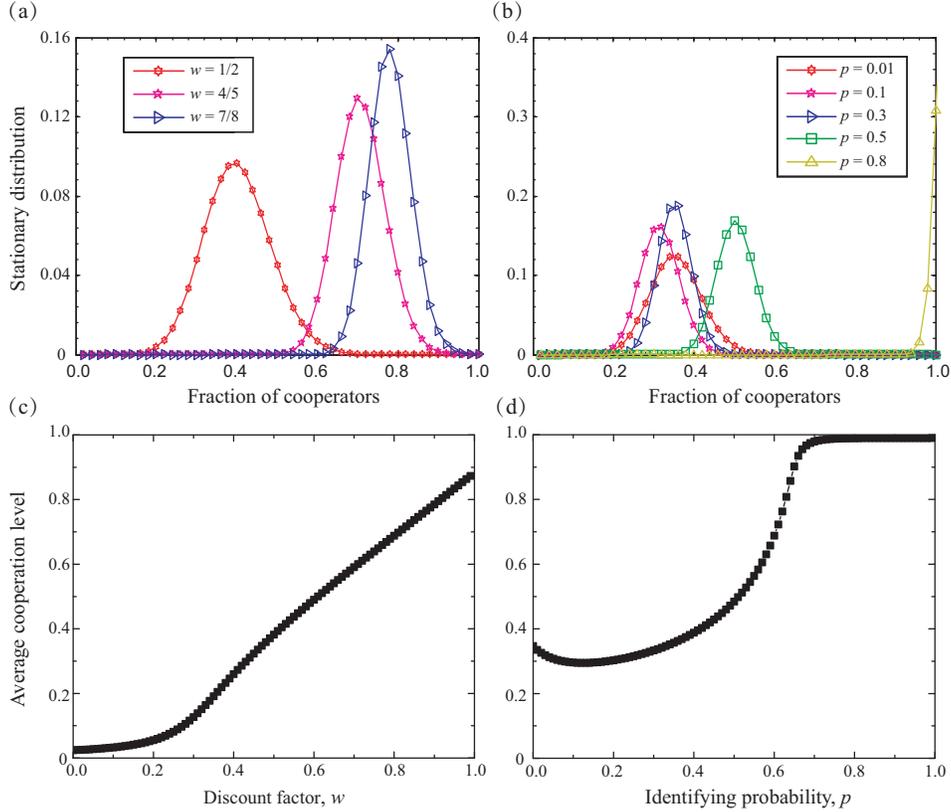}
\caption{The stationary distributions and the average level of cooperation change with model parameters. Top row shows that the stationary distributions of Marvok process vary with different values of $w$ and $p$. Bottom row respectively shows the average level of cooperation as functions of $w$ and $p$. Parameter values: $Z=50, N=5, c=1, \mu=0.01, s=2,$ and $p=0.6$ in panels (a) and (c); $Z=50, N=5, c=1, \mu=0.01, s=2,$ and $T=5$ in panels (b) and (d).}
\label{fig5}
\end{figure}

In the top row of figure~\ref{fig5}, we show that the stationary distributions of the system change with different model parameters ($w$ and $p$) when strategy mutation is considered. We find that the population spends most of the time in configurations where cooperators and defectors coexist for different $w$ values (see figure~\ref{fig5}(a)). Besides, with the increase of $w$, the population will spend a greater amount of the time in a more cooperative configuration. figure~\ref{fig5}(b) shows that the system
spends most of the time in the full cooperation state when $p$ is high. When the value of $p$ is lower than an intermediate value, we find that there exists a nonlinear phenomenon in the shape of the stationary distributions with the increase of $p$.

In the bottom panels of figure~\ref{fig5}, we respectively show how the average level of cooperation varies with model parameters $w$ and $p$. We find that the average level of cooperation increases with increasing $w$, which means that the larger the average number of game rounds, the easier it is to maintain a high level of cooperation (see figure~\ref{fig5}(c)). Besides, we observe that with the increase of $p$, the average level of cooperation first
decreases, and then increases until full cooperation is reached (see figure~\ref{fig5}(d)). Altogether, figure~\ref{fig5} reveals that cooperation can be better promoted when the discount factor $w$ is large or the identifying probability $p$ is high.

Finally, we stress that our main results are robust to other model parameters. Concretely, we find that similar evolutionary outcomes shown in infinite populations can still appear in a finite population for different values of identifying probability and discount factor (see figures~\ref{fig1}-\ref{fig4}). In addition, we investigate the effect of the group size $N$ on the evolutionary results both in infinite and finite populations and find that the main results in infinite and finite populations are robust when the group size changes appropriately. Besides, the cooperation cost, which is a key parameter characterizing the dilemma strength, plays an important role in the evolution of cooperation \cite{Tanimoto2007,wang2015,Arefin2020}. However, we find that our main results remain valid when the value of cooperation cost is
approximately changed.

\section{Conclusions}
Exclusion strategy has been regarded as an important incentive for the evolution of cooperation \cite{sasaki_prsb_2013,liu_ND_2019,quan_amc_2020}. In general, the exclusion is manipulated by refusing free-riders to be the beneficiaries
of the game, but at a cost to excluders. We say that this kind of exclusion is the direct one. In daily
interactions, exclusion may also appear in an indirect way, that is, the organizer tends to choose
more cooperative individuals to maintain their future interactions, while terminating upcoming
interactions with free-riders. At present, there is still no research to explore the evolutionary
dynamics of cooperation induced by indirect exclusion from a theoretical perspective.

In this paper, we have introduced indirect exclusion into the repeated PGG where the game
organizer probabilistically selects cooperators from the first game round to participate in the
subsequent group interactions and investigated its evolutionary dynamics in infinite and finite
well-mixed populations. Through theoretical analysis and numerical calculations, we have found
that the introduction of indirect exclusion can solve the cooperation conundrum in two forms:
first, cooperators and defectors can coexist stably in the population; second, cooperation is the
dominant strategy. Besides, we have shown that the identifying probability of the organizer has
a nonlinear effect on the collective behaviour when its value is not high, while higher values of
identification probability are helpful to the construction of a fully cooperative society. Our results
also indicate that more rounds of group interactions are conducive to maintaining a high level of
cooperation.

It is worth emphasizing that the second-order free-rider problem can be avoided in our
model, which is significantly different from the previous research about direct exclusion. Indeed,
previous studies have usually assumed that excluders need to pay an additional cost to expel
free-riders, thus cooperators (second-order free-riders) have an evolutionary advantage over excluders \cite{sasaki_prsb_2013,liu_jrsf_2022}. However, in our model, we do not assume an extra exclusion fund for
excluders, but alternatively consider that the game organizer probabilistically selects cooperators
to participate in the game interactions, which directly prevents the presence of the second-order
social dilemma.

The key feature of the presented model is the indirect approach for social exclusion, where a
game organizer is selected who has the power to select cooperators after the first game round, to
participate in the following interactions. A natural extension of the current model is to consider
that the cooperators' selection is performed not just after the first round, but several rounds \cite{liu_jrsf_2022}. Such consideration can help to build sufficient trust or reputation for individuals in the group,
especially when noise is non-negligible \cite{Han2021,Perret2021}. In this work, we have considered that the selection
of game organizer is random and has nothing to do with individual reputation. Therefore, it
is interesting to explore the role of indirect exclusion in the evolution of cooperation when the
game organizer is selected via reputation or based on past interactions \cite{Krellner2022}. Furthermore, as we emphasize in our model, we do not consider the strategy with memory or conditional strategy. Previous work has characterized partner strategies, competitive strategies, and zero-determinant strategies within the class of memory-one strategies for the iterated prisoner's dilemma, and presented some interesting properties of these strategies \cite{Hilbe2015}. It is thus worth exploring the effect of memory strategies with different properties on cooperation in the scenario of indirect exclusion in repeated group interactions.

\enlargethispage{20pt}


\section*{Acknowledgments}
This research was supported by the National Natural Science Foundation of China (Grants Nos. 61976048 and 62036002) and the Fundamental Research Funds of the Central Universities of China. L.L. acknowledges the support from Special Project of Scientific and Technological Innovation (Grant No. 2452022107).

\section*{Declaration of interest}
We declare we have no competing interests.




\end{document}